# BLIND CHANNEL ESTIMATION ENHANCEMENT FOR MIMO- OFDM SYSTEMS UNDER HIGH MOBILITY CONDITIONS

Aida Zaier[1] and Ridha Bouallègue[2]

[1]Department of Telecommunications, 6'Tel Research Unit, National Engineering School Tunis, Tunisia
`zaieraida@yahoo.fr`
[2] 6'Tel Research Unit, Higher School of Communications of Tunis, Tunisia
`ridha.bouallegue@gnet.tn`

## ABSTRACT

*In this paper, we propose an enhancement of a blind channel estimator based on a subspace approach in a MIMO OFDM context (Multi Input Multi Output Orthogonal Frequency Division Multiplexing) in high mobility scenario. As known, the combination between the MIMO context and the OFDM system has stimulated mainly the evolution of the fourth generation broadband wireless communications. The simulations results have demonstrated the effectiveness of the approach for a 16 QAM modulation scheme and had been evaluated in term of bit error rate BER and mean square error MSE versus the signal to noise ratio SNR.*

## 1. INTRODUCTION

Multiple-Input-Multiple-Output (MIMO) antennas with Orthogonal Frequency Division Multiplexing (OFDM) provide high data rates and are robust to multi-path delay in wireless communications [1]. Channel parameters are required for diversity combining, coherent detection and decoding. Therefore, channel estimation is essential in MIMO-OFDM system design.

The fundamental phenomenon which makes reliable wireless transmission difficult is the multipath fading. Hence, in ODFM Systems, there are two phenomena that can not be ignored: the distortion and the inter symbol interference (ISI) provoked by frequency selective fading channels. [2][3]

Thus, it appears clearly that a combination of both technologies, to obtain a MIMO-OFDM system is well promising for achieving higher data rate and larger system capacity over mobile wireless links recently.

In order to achieve these performance improvements, accurate CSI (Channel State Information) is required at the receiver which is obtained via channel estimation.

The most of channel estimation schemes was extended from SISO OFDM context to MIMO OFDM especially those based on training sequences (pilot or preamble)

MIMO-OFDM channel estimation techniques can be classified into the following three classes: training-based methods, blind methods and semi blind methods. The first one (training-based methods) employ known training signals to provide accurate channel estimation and thus involve consumption of spectral resources [4][5]., as examples of the algorithms used for this aim, we can cite: the least square (LS), maximum-likelihood (ML), and minimum mean square error (MMSE) methods. The second category concerns blind MIMO-OFDM channel estimation algorithms, which exploit the second-order stationary statistics, correlative coding, and other properties, normally have better spectral efficiency. With a small number of training symbols, the last class which is semi blind methods have been proposed to estimate the channel ambiguity matrix in MIMO-OFDM systems [6][7]. Note that most of the existing blind and semi blind methods for MIMO OFDM channel estimation, except for several algorithms that are proposed for orthogonal space–time-coded systems are based on the second-order statistics of a long vector, whose size is equal to or larger than the number of sub carriers.

A promising family of blind algorithms is the so-called subspace based blind channel estimation algorithms, which derive their properties from the second-order statistics of the received signals [8][9]. The subspace method has simple structure and achieves good performance, but there are two difficulties when applied to MIMO systems. Firstly, more receive antennas than transmit antennas are required to find noise subspace. Secondly, the precise knowledge of the channel order must be obtained, which is very difficult in practice. Several blind channel estimation algorithms for the single out single in (SISO) OFDM systems based on a non redundant linear precoder at the transmitter have been recently proposed in [10],[11] and [12]. In these methods, a non redundant linear precoder is applied, which can introduce a correlation structure in signals transmitted over different subcarriers and avoid the catastrophic effects of channel zeros at certain subcarriers. [8]

In this study, we have mainly enhanced the performances of the blind estimation of the channel based on a subspace approach. This estimator had demonstrated also a good behaviour in term of channel efficiency.

This paper is organized as follows. Section II provides a description of the MIMO system model. The next section presents the channel estimation scheme. Section IV covers the simulations results which are performed in terms of Bit Error Rate and Mean Square Error. Finally, the conclusion is specified in section V.

## 2. MIMO OFDM SYSTEM MODEL

The MIMO OFDM system model used for a blind based channel estimation scheme is illustrated with the figure below:

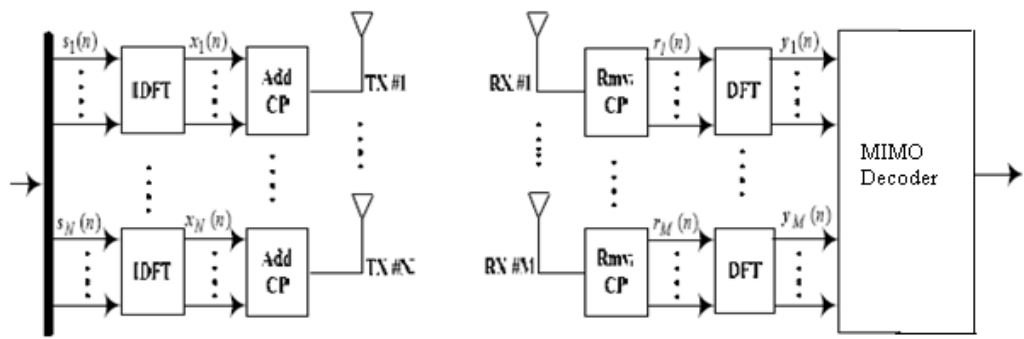

Figure 1: Transmitter and receiver structure of an $N \times M$ MIMO OFDM System [13]

We consider a MIMO-OFDM system with K subcarriers, N transmit and M receive antennas, signaling through frequency selective fading channel. Suppose all the $N \times M$ channel paths have the memory upper bounded by L, and let $h_{i,j} = [h_{i,j}(0), ..., h_{i,j}(L-1)]_{L \times 1}^T$ is the *L-th* order frequency selective fading channel between the *i-th* transmit antenna and the *j-th* receive antenna which is quasi-static, that is, it is fixed during the transmission of one packet and varies independently for each packet. [13]

Let us define the *n-th* OFDM block to be transmitted from *i-th* (*i*=1,…,*N*) transmit antenna as $s_i(n) = [s_i(0,n), ..., s_i(K-1,n)]_{K \times 1}^T$ where $s_i(k,n)$ is the *n-th* transmitted symbol from the *i-th* transmit antenna at the *k-th* subcarrier. The corresponding time-domain block is given by

$$x_i(n) = F^H s_i(n) \tag{1}$$

where $F$ is the *K*-point discrete Fourier transform (DFT) matrix with entries $F(a,b) = \frac{1}{\sqrt{K}} e^{-2j\pi(a-1)(b-1)/K}$. The CP is added at the front of each transmitted block and is removed at each received block. As long as the length of the CP is greater than or equal to *L*, the remaining signal at the *j-th* receive antenna for the *n-th* block can be expressed as:

$$r_j(n) = \sum_{i=1}^{N} \bar{H}_{i,j} F^H s_i(n) + \bar{v}_j(n), \qquad j = 1, ..., M \tag{2}$$

Where $\bar{v}_j(n)$ is the vector describing the unknown white Gaussian noise of dimension $K \times 1$ at the *j-th* receive antenna with equivalent variance $\sigma_{v_j}^2$ at each sampling time, and $\bar{H}_{i,j}$ is the $K \times K$ circulant channel matrix with its *(a,b)-th* entry given by $h_{i,j}^0([a,b] \bmod K)$ where $h_{i,j}^0$ is a vector obtained by adding $K - L - 1$ zeros at the end of $h_{i,j}$. The K point DFT of the channel matrix can be expressed as: $H_{i,j} = DFT(h_{i,j}) = [H_{i,j}(0), ..., H_{i,j}(K-1)]_{K \times 1}^T$

The normalized DFT of the received signal vector from the *jth* receive antenna is then represented as:

$$y_j(n) = F r_j(n) = \sum_{i=1}^{N} F \bar{H}_{i,j} F^H s_i(n) + F \bar{v}_j(n) \tag{3}$$

$$= \sum_{i=1}^{N} \widetilde{H}_{i,j} s_i(n) + v_j(n), j = 1, ..., M$$

Where $\widetilde{H}_{i,j} = diag(H_{i,j})$ is a diagonal matrix with the diagonal elements obtained from $H_{i,j}$. It can be easily shown that the new random noise vector $v_j(n)$ has the same statistical distribution as $\bar{v}_j(n)$

Then, the beyond system model can be rewritten by a matrix notation as:

$$\boldsymbol{y(n) = \mathcal{H}_y s(n) + v(n)} \tag{4}$$

Or in other representation as [4]:

$$\begin{bmatrix} y_1(n) \\ \vdots \\ y_M(n) \end{bmatrix}_{MK \times 1} = \begin{bmatrix} \widetilde{H}_{1,1} & \cdots & \widetilde{H}_{N,1} \\ \vdots & \ddots & \vdots \\ \widetilde{H}_{1,M} & \cdots & \widetilde{H}_{N,M} \end{bmatrix}_{MK \times NK} \begin{bmatrix} s_1(n) \\ \vdots \\ s_N(n) \end{bmatrix}_{NK \times 1} + \begin{bmatrix} v_1(n) \\ \vdots \\ v_M(n) \end{bmatrix}_{MK \times 1} \tag{5}$$

In (5), $\mathcal{H}_y$ denotes the overall frequency-domain channel matrix with its *(j,i)-th* partitioned block given by $\widetilde{H}_{i,j}$

The source covariance matrix is defined as:

$$R_s = E[s(n)s^H(n)] = \sigma_s^2 I_{KN} \qquad (6)$$

Where $\sigma_s^2$ is the transmitted signal power.

## 3. BLIND CHANNEL ESTIMATOR

The input-output relationship for a MIMO system with N transmit and M receive antennas and frequency selective finite impulse response (FIR) multipath channel. Assuming that the total bandwidth is divided into K orthogonal tones. At transmission time n, the complex baseband data blocks, which are modulated by linear modulator, such as M-PSK or M-QAM, are expressed as [14]:

$$d_i(n) = (d_i[n,0], \quad d_i[n,1], \ldots, d_i[n,K-1])^T, i = 1,2,\ldots,M \qquad (7)$$

These data blocks forms OFDM blocks which will reach the OFDM modulator which in its turn applies a K-point IFFT to every OFDM block and prepends the cyclic prefix of length P. Then the resulting data blocks are expressed as

$$s_i(n) = (s_i[n,K-1], \quad s_i[n,K-2], \ldots, s_i[n,0], \ldots, s_i[n,P])^T, i = 1,2,\ldots,M \qquad (8)$$

Assume that $W_K = \exp(j\frac{2\pi}{K})$, $\tilde{W} = I_J \otimes W$ with $\otimes$ is the Kronecker product, the relation between s and d become:

$$s = \tilde{W}d \qquad (9)$$

In a notation matrix, we write the received signal as follow:

$$r = Hs + b = H\tilde{W}d + b = Ad + b \qquad (10)$$

After setting the cyclostationarity conditions of the channel outputs given in [15] and establish the sufficient identifiability conditions, we can write the autocorrelation matrix $R_i$ of the received data vector r as:

$$R_r = E[rr^H] \qquad (11)$$

The noise is assumed to be independent of the transmitted sequence. Then $R_r$ is expressed as

$$R_r = AR_d A^H + R_b \qquad (12)$$

Where

$$R_b = E[bb^H] = \sigma_b^2 I_{J(K+P)N-LN}$$

$$R_d = E[dd^H]$$

$R_d$ is assumed to be full rank. Since the channel noise is assumed to be AWGN, The eigenvalue decomposition of $R_r$ has the form of:

$$R_r = U\, diag(\lambda_1 \ldots \lambda_{JKM} \quad \lambda_{JKM+1} \ldots \lambda_{J(K+P)N-LN})U^H \qquad (13)$$

where the columns of U are eigenvectors, $\Lambda$, are the eigenvalues, and $\lambda_1 \geq \lambda_2 \geq \cdots \geq \lambda_{JKM} \geq \cdots \geq \lambda_{J(K+P)N-LN}$ [27]. Denote the eigenvectors associated with the eigenvalues $\lambda_{JKM}\ \lambda_{JKM+1}\ \ldots\ \lambda_{J(K+P)N-LN}$ by $G_1, \ldots, G_g$ and $G = [G_1\ \ldots\ G_g]$ where $g = JK(N-M) + JPN - LN$

Then G spans the null space of $AR_d A^H$ and is orthogonal to its range space. Thus, we have:

$$G_i^H A = 0 \quad i = 1,2,\ldots,g \tag{14}$$

Then the channel estimator is given from an optimizing quadratic cost function as:

$$\hat{h} = \underset{\|h\|=1}{\operatorname{argmin}} \left\{ \sum_{i=1}^{g} G_i^H A A^H G_i \right\} \tag{15}$$

## 4. SIMULATIONS RESULTS

In this section, we will present the main results to demonstrate the effectiveness of the proposed scheme.
The channel is a Rayleigh fading one. The performance of the blind channel estimation is evaluated with the mean square error (MSE) of the estimated channel coefficients and also in term of the bit error rate (BER) versus the signal to noise ratio (SNR).
The simulations were run for a MIMO OFDM with 4×4 scheme and for a 16 QAM modulation.

The figure 2 below shows the real and the imaginary part of the channel. It illustrates the estimate and the accurate part of the channel.

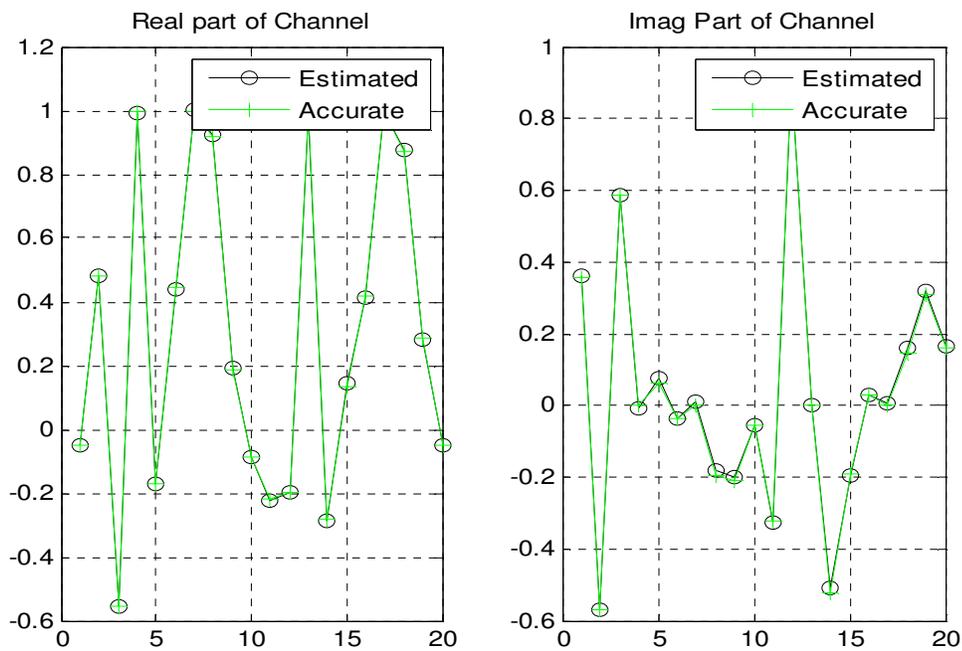

Figure 2: Estimate and exact components of the real and imaginary parts of the channel

International Journal of Wireless & Mobile Networks (IJWMN) Vol. 4, No. 1, February 2012

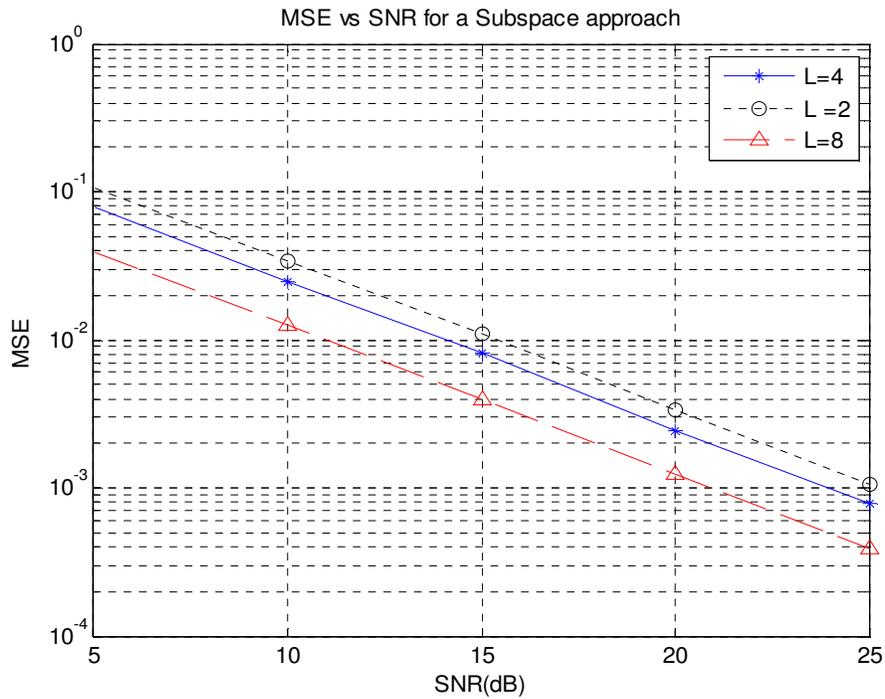

Figure 3: MSE vs SNR for a subspace approach with different length of channel

This figure shows clearly that we have obtained better values comparably with related works and the mean square error had been mainly improved.

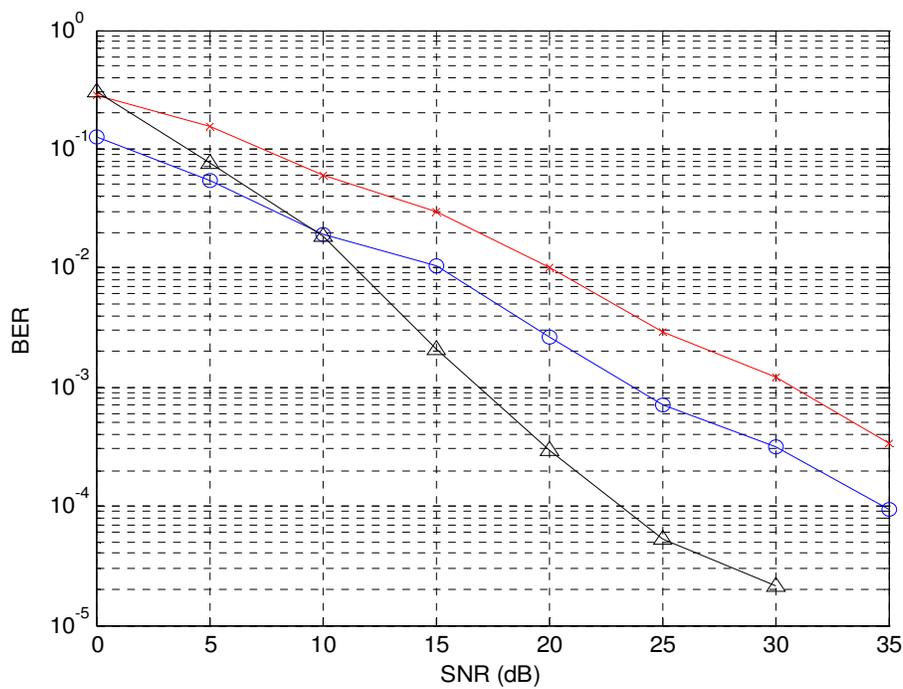

Figure 4: BER vs SNR with different length of the channels

The figure above illustrates the BER for various values of the length of the channels and with different Doppler spread. We can judge that the results are satisfying and are enhanced comparing with other works. Under 10 dB we see that the BER is the lower for a channel length of 4 but is will be better for a length of 8 when the SNR exceeds 10 dB.

## 5. CONCLUSIONS

In this paper, we presented an enhancement of the blind channel estimation in multi-antenna OFDM systems operating over frequency selective fading channels. The subspace approach, already showed its efficiency and flexibility compared to the other channel estimation schemes.
The performance of the enhanced estimator was investigated by computer simulations. The results clearly show the effectiveness of our proposed algorithm, as well as its improvement over the existing subspace-based techniques.
Theoretical analysis and simulation results show that our algorithm has a good performance for high Doppler spread.

**Authors**

**PR. RIDHA BOUALLEGUE**
Received the Ph.D degrees in electronic engineering from the National Engineering School of Tunis. In Mars 2003, he received the Hd.R degrees in multiuser detection in wireless communications. From September 1990
He was a graduate Professor in the higher school of communications of Tunis (SUP'COM), he has taught courses in communications and electronics. From 2005 to 2008, he was the Director of the National engineering school of Sousse. In 2006, he was a member of the national committee of science technology. Since 2005, he was the laboratory research in telecommunication Director's at SUP'COM.
From 2005, he served as a member of the scientific committee of validation of thesis and Hd.R in the higher engineering school of Tunis. His recent research interests focus on mobile and wireless communications, OFDM, OFDMA, Long Term Evolution (LTE) Systems. He's interested also in space-time processing for wireless systems and CDMA systems.

**ZAIER AIDA**

Received the B.S. degree in 2005 from National Engineering School of Gabes, Tunisia, and M.S. degree in 2006 from Polytechnic School of Sophia Antipolis of Nice Frrance. Her Research interests focus on channel estimation and synchronization of OFDM and MIMO-OFDM channels under very high mobility conditions.